\documentclass[aip,apl,twocolumn,tightenlines,superscriptaddress,10pt]{revtex4}
\usepackage{graphics,bm}
\usepackage{amssymb}
\usepackage{amsmath}
\usepackage{epsfig}
\usepackage{epsf}
\usepackage[usenames]{color}

\begin{document}
\def\m{{\bf m}}
\def\s{{\rm s}}
\def\sd{{\rm sd}}
\def\F{{\bf F}}
\def\nabvec{{\boldsymbol{\nabla}}}
\def\P{{\mathcal P}}
\def\l{{\ell}}
\def\x{\vec{x}}
\def\ii{{\hat\i}}
\def\jj{{\hat\j}}
\def\om{{\bf \Omega}}
\def\b{\hat{b}}
\def\bd{\hat{b}^\dag}
\def\k{\vec{k}}
\def\q{\vec{q}}
\def\ad{a^\dag}
\def\A{\vec A}
\def\x{\vec{x}}
\def\bx{{\bf x}}
\def\bk{{\bf k}}

\title{Spin-transfer mechanism for magnon-drag thermopower}

\author{M. E. Lucassen}
\email{m.e.lucassen@uu.nl}
\author{C. H. Wong}
\author{R. A. Duine}
\affiliation{Institute for Theoretical Physics, Utrecht
University, Leuvenlaan 4, 3584 CE Utrecht, The Netherlands}

\author{Y. Tserkovnyak}
\affiliation{Department of Physics and Astronomy, University of California, Los Angeles, California 90095, USA}

\begin{abstract}
We point out a relation between the dissipative spin-transfer-torque parameter $\beta$ and the contribution of magnon drag to the thermoelectric power in conducting ferromagnets. Using this result we estimate $\beta$ in iron at low temperatures, where magnon drag is believed to be the dominant contribution to the thermopower. Our results may be used to determine $\beta$ from magnon-drag-thermopower experiments, or, conversely, to infer the strength of magnon drag  via experiments on spin transfer.
\end{abstract}

\maketitle

A recurring theme in the field of spintronics is the interplay between electric and spin currents, and magnetization dynamics in conducting ferromagnets. This activity was initiated by the theoretical prediction of Slonczewski \cite{slonczewski1996} and Berger \cite{berger1996} who showed that magnetic layers in nanopillars can be excited or even reversed by spin-polarized currents \cite{tsoi1998}. The underlying mechanism is dubbed spin transfer, as it involves the transfer of spin angular momentum from conduction electrons to magnetization. In part because of its promise for applications such as magnetic memories, spin transfer is now actively studied in the context of current-driven domain wall motion in magnetic nanowires \cite{grollier2003}. As a result of these efforts, it is now understood \cite{tataraPRP06} that there are at least two contribution to spin transfer in the long-wavelength limit, one reactive (sometimes called adiabatic) \cite{bazaliyPRB98} and one dissipative \cite{zhang2004}. This latter torque is parameterized by a dimensionless constant $\beta$, and the ratio of this constant to the Gilbert magnetization damping constant $\alpha$ is of crucial importance for the phenomenology of current-driven domain-wall motion \cite{thiaville2005,tserkovnyak2006}. Precise experimental determination of $\beta$ from domain-wall experiments \cite{heyne2008} or experiments on magnetic vortices \cite{bolte2008} is, however, difficult.

A closely-related development is the study of spin and charge currents induced by time-dependent magnetization, called spin pumping in layered systems \cite{tserkovnyak2002}, and usually referred to as spin motive forces in magnetic textures \cite{barnes2007}. The latter were observed in a very recent experiment on field-driven domain walls \cite{yang2009}, and proposed for magnetic vortices \cite{ohe2009}. Like spin transfer, spin motive forces have two contributions corresponding to the reciprocal of the reactive and dissipative spin-transfer torques \cite{duine2008}. In particular, the current induced by a time-dependent magnetization texture also depends on the parameter $\beta$.

In this Letter, we show that $\beta$ is determined by the thermoelectric power due to electron-magnon scattering, the so-called magnon-drag thermopower \cite{bailyn1962}. This result is derived by considering the electric current density induced by a time-dependent magnetization, with direction determined by the unit vector $\m (\x,t)$, which is given by \cite{barnes2007,duine2008}
\begin{align}\label{eq:electromotive}
 \vec j=- \frac{\hbar P\sigma }{2|e|}\left[ {\bf m}  \cdot \left(\frac{\partial{{\bf m} }}{\partial t} \times  \frac{\partial{{\bf m} }}{\partial \x} \right)+\beta \frac{\partial{{\bf m} }}{\partial t} \cdot\frac{\partial{{\bf m}}}{\partial \x}\right]~,
\end{align}
where $-|e|$ is the electron charge, $\sigma$ the electrical conductivity, and $P$ is the current spin polarization. Although the above expression is usually considered for magnetization textures such as domain walls or magnetic vortices, it is straightforwardly evaluated for a magnetic configuration corresponding to a transport steady state of (Holstein-Primakoff) magnons. This results in \cite{footnote}
\begin{align}\label{eq:currentrelation}
\vec j=\beta \frac{\hbar^2 \gamma P\sigma}{2 |e|M_s D}\vec j_{Q,m}~,
\end{align}
where $M_s$ is the saturation magnetization density, and $\gamma$ is the (minus) gyromagnetic ratio. Equation \eqref{eq:currentrelation} shows that a magnon heat current $\vec j_{Q,m}$ results in an electrical current $\vec j$. The main assumption leading to the above result is that the energy of magnons with wave vector $\vec k$ is equal to $\hbar \omega_{\vec k} = D k^2$, in terms of the spin stiffness $D$. This is a valid approximation for temperatures  larger than the magnon gap, which is typically $\sim 1$ K in metallic ferromagnets.

To understand how the above result is related to magnon-drag thermopower, we consider the response of the system to electric field $\vec E$, magnon-temperature and electron-phonon-temperature gradients, denoted by $\vec \nabla T_m$ and $\vec \nabla T_{e,p}$, respectively. Introducing two different temperatures for these subsystems is in the present case needed to make connection with the result in Eq.~(\ref{eq:currentrelation}). We note that in theoretical discussions \cite{xiao2010} of the spin-Seebeck effect \cite{uchida2008} such temperature differences are also invoked. The linear-response coefficients are determined by
\begin{align}
\begin{pmatrix}\vec j\\\\ \vec j_{Q}\\\\ \vec j_{Q,m}\end{pmatrix}=\begin{pmatrix}\sigma&\sigma S_{e,p}T &\sigma S_{m} T\\\\
\sigma S_{e,p}T &\kappa'_{e,p} T& \zeta T \\\\
\sigma S_{ m}T & \zeta T &\kappa'_{m}T
\end{pmatrix}\begin{pmatrix}\vec E\\\\
- \frac{\vec \nabla T_{e,p}}{T}\\\\ -\frac{\vec\nabla T_{m}}{T}\end{pmatrix}\;,
\label{response}
\end{align}
where $\vec j_Q$ is the heat current carried by electrons and phonons. In the above, the magnon-drag thermopower is denoted by $S_m$, and the magnon heat conductivity at zero electric field by $\kappa_m'$. The contribution of electrons and phonons to the thermopower is denoted by the Seebeck coefficient $S_{e,p}$, and their heat conductivity at zero field by $\kappa'_{e,p}$. Drag effects between magnon heat currents and electron-phonon heat currents are denoted by $\zeta$. Also note that we have used Onsager relations to eliminate the Peltier coefficients.

It is important to point out that disentangling heat currents in the above way only applies to weakly-coupled situations. In case this is not possible, such that only the total Seebeck coefficient and thermal conductivity can be measured, our results below are applicable to the case that the thermal transport is dominated by magnons.

The result in Eq.~(\ref{eq:currentrelation}) applies to the situation that the electron-phonon temperature gradient and electric field are zero. Taking $\vec E=\vec \nabla T_{e,p}=0$  and $\vec \nabla T_m \neq 0$, we find a magnon heat current and a charge current that are proportional to $\vec \nabla T_{m}$ such that we have $\vec j=\vec j_{Q,m} \sigma S_{m}/\kappa'_{m}$. We combine this with Eq.~(\ref{eq:currentrelation}) to find our main result
\begin{align}\label{eq:betarelation}
\beta=\frac{2 |e|M_s D}{ \hbar^2 \gamma P}\frac{S_{ m}}{\kappa'_{ m}}\;.
\end{align}
This result relates the spin-torque parameter $\beta$ to the magnon-drag thermopower and the magnon heat conductivity at zero field $\kappa'_m$. The magnon heat conductivity $\kappa_m$ at zero electric current, defined by $j_{Q,m}=-\kappa_m\nabla T$ with $T_m=T_{e,p}=T$, in terms of the above transport coefficients, is given by $\kappa_m = \kappa'_{ m}+\zeta-\sigma S_{ m}(S_{ e,p}+S_{ m})T$. The last correction is small in most materials, except for very good thermoelectric materials. Assuming that the magnon-electron heat drag is small, i.e., $\zeta\ll\kappa'_{ m}$, we take $\kappa_{ m}'\approx \kappa_{ m}$ in our estimates.

We now estimate $\beta$ using available experimental data on magnon-drag thermopower and magnon heat conductivity. In this order-of-magnitude estimate we take, for simplicity, $P=1$, $\gamma=2 \mu_B/\hbar$ ($\mu_B$ is the Bohr magneton) and $M_s=\mu_B/a^3$ with $a\simeq 0.3$ nm a typical lattice constant. According to Blatt {\it et al.} \cite{blatt1967}, the main contribution to the thermopower in iron at low temperatures is due to magnon drag and they give the result $S_m \approx 0.016~(T/K)^{3/2}~\mu$V$/$K. Hsu and Berger \cite{hsu1978} find the value of $\kappa_m=4.9 \times 10^{-2}$ W$/$K\,m for Fe$_{95}$Si$_5$ at $4$ K (the iron is silicon doped to decrease the electronic contribution to the heat conductivity). Using a typical value $D=4 \times 10^{-40}$ J\,m$^2$ \cite{hsu1978} for the spin stiffness, we find that $\beta\approx 0.1$ at $4$ K. The main uncertainty in our estimate is the value of $\kappa_m$ which is difficult to measure. Nonetheless, this value for $\beta$ seems not unreasonable as room-temperature values for this parameter obtained from spin-transfer experiments usually  find that $\beta \sim 0.1-0.01$ for Permalloy \cite{heyne2008,bolte2008}. We also point out that the $T^{3/2}$ temperature scaling of $S_m$ would imply, according to Eq.~\eqref{eq:betarelation}, that $\kappa_m\propto T^{3/2}/\beta$. It can be shown, on the other hand, that $\kappa_m\propto T^{3/2}/\alpha$ \cite{kovalevCM11} within the Landau-Lifshitz-Gilbert phenomenology, suggesting the ratio $\beta/\alpha$ is insensitive to temperature, which, in turn, is supported by microscopic calculations \cite{tserkovnyak2006,kohno2006}.

The transport coefficients in Eq.\,\eqref{response} determine the dissipation, which must be positive by the second law of thermodynamics.   This imposes the condition that the determinant of the response matrix be positive, which is satisfied if
\begin{equation}
 T \sigma\left( \frac{S_{e,p}^2}{\kappa'_{e,p}}+\frac{S_m^2}{\kappa'_m}\right)
+\frac{\zeta}{\kappa'_{e,p}\kappa'_{m}}\left(\zeta-2 S_{e,p}S_m \right)\leq1\; .
\end{equation}
It is conventional to define $Z'_{e,p}=\sigma S_{e,p}^2/\kappa'_{e,p}$,  $Z'_{m}=\sigma S_{m}^2/\kappa'_{m}$, so that this relation reads $Z'_{e,p}T+Z'_{m}T\leq1$, where we assumed $\zeta\ll\kappa'_{ m}$ like before.  Using Eq. \eqref{eq:betarelation}, this condition imposes an upper bound on $\beta$:
\begin{equation}
\beta\leq\frac{2 |e|M_s D}{ \hbar^2 \gamma P}\frac{1}{\sigma  S_m T}\left(1-Z'_{e,p} T\right)\;,
\end{equation}
where $\sigma S_m$ and $\gamma P $ are assumed positive, which is typically the case. Using the same values as in our previous estimate and taking $Z'_{e,p} T \ll 1$, we find that $\beta  \lesssim 1$ at $4$ K, using a value of $\sigma \approx 10^{11}$ $\Omega$\,m at $4$ K \cite{kemp1956}. This result gives an upper bound for $\beta$ for a material, which is particularly useful when looking for materials with large $\beta$. This is of particular interest for spintronics applications, since large $\beta$ implies a large current-to-domain-wall coupling.

In conclusion, we have shown that the spin-transfer-torque parameter $\beta$ is related to the ratio of the magnon-drag thermopower and the magnon heat conductivity. From an experimental point-of-view this relation can be used to either determine $\beta$ experimentally, or to obtain information on the contribution of magnons to heat conduction and thermopower from experimental knowledge of $\beta$. From a theoretical point-of-view the relation derived in this letter opens the way for new methods to calculate $\beta$. The microscopic calculations in the literature usually focus on the contribution to $\beta$ due to spin-dependent disorder scattering \cite{tserkovnyak2006,kohno2006} or take into account scattering phenomenologically \cite{garate2009}. In future work we intend to microscopically determine $\beta$ by calculating the magnon-drag thermopower and magnon heat conductivity and then using the relation in Eq.~(\ref{eq:betarelation}). We expect that this approach will be particularly useful in determining the temperature dependence of $\beta$.

This work was supported by Stichting voor Fundamenteel Onderzoek der Materie (FOM), the Netherlands Organization for Scientific
Research (NWO), by the European Research Council (ERC) under the Seventh Framework Program (FP7), the Alfred P. Sloan Foundation, and by the NSF under Grant No. DMR-0840965.

\end{document}